# Thunderstorm Ground Enhancements: Multivariate analysis of 12 years of observations

A. Chilingarian[1,2], G. Hovsepyan[1], D. Aslanyan[1], T. Karapetyan[1], Y. Khanikyanc[1], L.Kozliner[1], D. Pokhsraryan[1], B. Sargsyan[1], S.Soghomonyan[1], S.Chilingaryan[1], and M.Zazyan[1]

[1]*A. Alikhanyan National Lab (Yerevan Physics Institute), Yerevan 0036, Armenia*
[2]*National Research Nuclear University MEPhI, Moscow 115409, RF*

**Abstract**

We present a survey of more than half-thousand thunderstorm ground enhancements (TGEs, fluxes of electrons, and gamma rays associated with thunderstorms) registered in 2008-2022 at Aragats space environmental center (ASEC). We analyze correlations between various measured parameters characterizing TGEs measured on Aragats. Two special cases of TGE events are considered: one, terminated by nearby lightning flashes, and another one – with a sufficiently large ratio of electrons to gamma rays. On the basis of the analysis, we summarize the most important results obtained during 12 years of TGE study, which include:
• We show the operation of the electron accelerators in the thunderous atmosphere by directly measuring the electron flux during thunderstorms;
• Quite frequently, TGEs occur prior to lightning flashes and are terminated by them.
• The energy spectra of avalanche electrons observed on Aragats indicate that the strong electric field region can extend very low above the ground covering a large area on the earth's surface;
• TGEs originated from multiple relativistic runaway electron avalanches (RREAs) starting with seed electrons from the ambient population of cosmic rays, which enter an extended region of the electric field with strength exceeding the critical value.

## 1. Introduction

The inventor of one of the first elementary particle detectors, Nobel laureate Sir Charles Thomson Rees Wilson nearly 100 years ago realized that in a strong accelerating electric field, collisions of electrons with atoms of the atmosphere do not prevent electrons from the acquisition of large kinetic energy [1, 2]. The results of the first experiments to observe the runaway electrons on the Earth's surface carried out by Schonland, and Halliday in South Africa using electroscopes, cloud chambers, and Geiger counters, were disappointing. The search for electrons with energies up to 5 GeV falling on the Earth's surface along the lines of force of the geomagnetic field, of course, could not give a positive result [3]. The geomagnetic field could only bend the trajectory of upward charged particles, but not in any way turn them towards the Earth. Wilson's idea that accelerated electrons can reach the atmosphere was confirmed after the launch of orbiting gamma-ray observatories. Numerous terrestrial gamma-ray flashes (TGFs, [4]) are observed at altitudes of ≈500 km above the Earth in correlation with strong equatorial thunderstorms. Of course, to detect the gamma-ray flash, the satellite must be in the right place (in a window of a few hundred km) and at the right time (in a window of a few milliseconds). The study of runaway electrons from thunderstorms with ground-based detectors is akin to the study of relativistic electrons produced in particle accelerators. The main difference is that the beam energy is not fixed and the beam cross-section can reach a few square kilometers. However, by placing particle detectors on the earth's surface in places where thunderclouds fall low millions of runaway avalanches can be observed.

After the first unsuccessful attempts to register particles coming from thunderclouds in South Africa in the thirties', in 1967, a 10% increase in 1-minute counts was recorded at the summit of Mount Lemmon (altitude 2800 m) at the Lighting Research Center of the University of Arizona [5]. The average duration of the surplus flux was 10 minutes; the threshold energy of the detector was 100 keV. Afterward, measurements carried out by an Italian group using the EAS-TOP ground complex [6] revealed a significant excess of the extensive air shower (EAS) count rate during a thunderstorm lasting 10–20 minutes. At a balloon flight altitude of 4 km, an increase in X-ray radiation (> 12 keV) above the background by a factor of ~ 3 was observed within 20 s [7]. A significant increase in X-ray fluxes at an altitude of 9 km, lasting about 10 s, and 10–100 times higher than the background flux was reported in [8]. There was also the first report of lightning abruptly stopping the particle's flux. In the 1990s, gamma-ray bursts (30–120 keV), 10–100 times higher than the background level, were recorded over Oklahoma at an altitude of 15 km (well above the upper limit of clouds) [9].

Recently, gamma-ray bursts have also been observed from aircraft. Over 37 flight hours at a cruising altitude of 14 to 15 km, 12 gamma-rays with energies from 50 keV to 5 MeV were recorded [10]. Their duration ranged from 4 to 112 s, and the brightest glow was 1-2 orders of magnitude higher than the background level and was suddenly stopped by lightning. In [11], two bursts of gamma radiation (from 100 keV to 10 MeV) were reported during a flight inside a thundercloud at an altitude of 12 km. Both luminescence lasted ~ 30 s, and the fluxes of gamma quanta exceeded the background level by 20 and 3 times, respectively. The first glow was suddenly stopped by the lightning flash. The above measurements were carried out mainly using spectrometers with small (3-7 cm) cylindrical crystals of sodium iodide activated by thallium (NaI (Tl)). A flux of high-energy gamma rays with a continuous energy spectrum up to 10 MeV, lasting up to 20 minutes, was recorded at the top of Mount Fuji (3776 m). The authors of [12] argue that bremsstrahlung photons generated by energetic electrons were born continuously due to a strong electric field in a thundercloud, and did not arise during a lightning discharge. At the Norikura research station in the Japanese Alps (2770 m above sea level), an increase in the flux of gamma rays and electrons was found during thunderstorms [13]. Since 2002, radiation monitoring services at nuclear power plants in Japan have reported gamma-ray bursts lasting about 1 minute, which correlate with thunderstorm activity. Gamma-ray bursts have been discovered during winter thunderstorms in Japan when thunderstorm clouds descend several hundred meters [14]; the increase of the natural gamma radiation reaches 50-100% [15,16]. In experiments at the Baksan Neutrino Observatory of the Institute for Nuclear Research of the Russian Academy of Sciences, time series of cosmic rays are continuously measured, as well as accurate measurements of the electric field and monitoring of thunderstorms are performed [17]. Changes in the intensity of soft cosmic rays (below 30 MeV) and hard cosmic rays (> 100 MeV) were studied. The ground detectors of the Baksan station are located in the gorge, so the clouds are located quite high, and the excess is usually several percent. The network of NaI detectors, together with the EAS launch system, is located at the Tien Shan cosmic ray station of the P.N. Lebedev Physical Institute of the Russian Academy of Sciences at an altitude of 3340 m. The aim of the study was to detect a runaway breakdown initiated by an EAS with an energy of more than 1000 TeV – so-called, RB-EAS discharge. (a discharge "generated in thunderclouds by joint action of runaway breakdown and extensive atmospheric shower" [18]).

Based on the short gamma-ray bursts (less than 200 seconds) detected by the network of gamma detectors, the authors of [19] argue that RB-EAS is a rather rare event (≈1% of all EAS recorded during thunderstorms), requiring several conditions to be met, the most important of which is the location of a strong electric field no higher than 300-400 m above the detector.

The theoretical explanation of electron multiplication and acceleration in thunderclouds was given by Alex Gurevich. He started to develop the theory of the electron avalanche as early as 1961 [20]. It was shown [21] that an energetic seed electron injected into a strong electric field creates an electron-photon avalanche that can reach the Earth's surface. A new physical phenomenon, which he called runaway breakdown, later was mostly cited as relativistic runaway electron avalanche (RREA, [22,23]). The first, and yet only observation of RREA electron flux and recovery of electron energy spectrum was made on Aragats [24]; electron-gamma ray avalanches covering thousands of square meters on the earth's surface also were observed on Aragats [25, 26].

## 2. Study of runaway avalanches at Aragats

From the brief overview given above, it is obvious that usually only one of the secondary types of cosmic rays was measured, and the number of detected "events with thunderstorm particles" was very modest. The available experimental data were insufficient confirm the theory of relativistic runaway avalanches, construct models of electron acceleration in thunderclouds, and to explain the charge structure of the cloud in which the runaway avalanche developed. What was needed, is multi-sensor ground-based observations using various particle detectors, systematically measuring fluxes of gamma rays, electrons, muons, and neutrons of atmospheric origin, their energy spectra, and correlations with the near-surface electric field (NSEF) and with lightning occurrences. These measurements should also be supplemented by meteorological and optical observations. In addition, it is highly desirable to have several remote observation sites for studies of the spatial structure of avalanches.

One of the world's largest high-altitude cosmic-ray research stations is located on Mt. Aragats 3200 m above sea level, near the southern peak of ≈2900 m on the beach of the large Kari Lake. Since 1942 physicists were studying Cosmic Ray fluxes on Mt. Aragats with various particle detectors: mass spectrometers, calorimeters, transition radiation detectors, and huge particle detector arrays detecting protons and nuclei accelerated in the most violent explosions in Galaxy. The latest research at Mt. Aragats include Space Weather, Solar accelerators, and high-energy physics in the atmosphere research with networks of particle detectors located on the slopes of Aragats in Armenia and abroad. From 2008 to 2022 particle detectors of the Aragats space environmental center (ASEC [27,28]) continuously registered fluxes of charged and neutral particles using various particle detectors located on the slopes of Aragats. ASEC detectors measure particle fluxes with different energy thresholds, as well as EASs initiated by primary protons and nuclei with energies above 50–100 TeV. Thunderstorm activity on Aragats is extremely strong in May - June. Sometimes lightning flashes occur in the close vicinity of the station, and lightning activity lasts for an hour or more. Thunderclouds are usually found below the southern peak (i.e., not higher than 500 m above ground level) and sometimes only 25–50 m above the station. Therefore, most energetic TGEs can contain a large portion of avalanche electrons, like one that occurred on 19 September 2009, which also triggers the MAKET surface array [29] proving that avalanche electrons are distributed over thousands of square meters of area [26].

Most of the particle detectors in 2009 were located in the MAKET experimental hall. First of all, it is the Aragats Solar Neutron Telescope (ASNT), which still remains the main detector in high-energy atmospheric physics, measuring the flux of electrons and gamma rays in the energy range 10-100 MeV [30]. In the same hall are located Aragats Neutron Monitor (ArNM), type 18HM64 and the space environment viewing and analysis network (SEVAN) particle detector, which records the fluxes of charged and neutral particles. 16 plastic scintillators of the MAKET-ANI surface array record both EASs and avalanches unleashed by runaway electrons originating in the thunderclouds above.

After the first years of TGE research, the experimental complex at Aragats was significantly expanded. Numerous new particle detectors were installed at an altitude of 3200 m. A network of 7 spectrometers (based on NaI crystals of 12 x 12 x 24 cm size was installed in the SKL experimental hall directly under a 0.6 mm thick iron roof. The low energy threshold (~ 300 keV) provides large statistics (~ 50,000 counts per minute) for reconstructing gamma ray differential energy spectra from 0.3 MeV. Detailed characteristics of the network of NaI detectors are given in Ref. [31]. A network of 3 STAND1 detectors (three stacked scintillators with a thickness of 1 cm and an area of 1 $m^2$ and one stand-alone with a thickness of 3 cm) is located at Aragats station premises with a spacing of ≈300 m. The network is connected to the fast data synchronization system that can capture time series with a sampling time of 50 ms, which allows one to study the relation of the TGE development to atmospheric discharges.

The largest TGEs registered by SEVAN network [32] units at Mt. Musala (Bulgaria) and Mt. Lomnicky Stit (Slovakia [33]), as well as the results obtained by the Japanese group [34], prove that TGE isn't only a specific Aragats feature, but – a universal characteristic of thunderstorms. The measured energy spectra allow us to get insight into the charge structure of the thundercloud and clarify the role of the lower positively charged region (LPCR) in the development of the TGE.

Data from local and international networks is transferred to the MySQL database at the CRD headquarters in Yerevan and is available through the ADEI multidimensional visualization and statistical analysis platform [35]. ADEI allows users to quickly analyze data, prepare figures and slides and perform joint data analysis with remote groups, test hypotheses, and draw physical inferences. Alerts and forewarnings sent by e-mail allow making it possible to follow the progress of thunderstorm events in real time. ADEI database contains time series of count rates of neutral and charged particles together with data on disturbances of the NSEF measured by a network of Boltek EFM-100 electric field mills and meteorological conditions from automatic weather stations from Davis Instruments. Placing these data in one database provides the possibility of the visualization and multivariate correlation analysis of particle fluxes and numerous environmental parameters.

Measurements of TGEs were supported by the simulation of the RREA development in the atmospheric electric field above detectors with GEANT4 [36] and CORSIKA [37] codes and spectrometric measurements of Radon progeny gamma radiation. Comparative analysis of observed TGEs and Radon isotope spectrograms allows explaining the shape of a long-lasting TGE [38], consisting of:

1. RREA phase. Large, reaching hundred percent flux enhancements, lasting a few minutes with particle energies reaching tens of MeV; fluxes are usually interrupted by a lightning flash. Particles come from the near-vertical direction.
2. Radon progeny radiation. Low-energy (<3 MeV) radiation is never interrupted by lightning; particle flux is isotropic.
3. The decay phase. Decay of radon progeny that is still concentrated in the air after the storm finishes. The half-life time of TGE decay is consistent with the half-life time of $^{214}$Pb (~300 keV peak) and $^{214}$Bi (~600 keV peak) isotopes from the radon chain.

Afterward, we considered scenarios of TGE development in the lower dipole. The scenarios of the origination of the downward electron-accelerating electric field are numerous, and corresponding TGEs may vary in intensity and energy spectra, as well as in the fraction of particles reaching Earth's surface. However, by making multiple simulations with different electric field strength and elongation we outline plausible parameters of the electric field that give rise to TGEs and define the main structures of the thundercloud charge structures supporting the origination of TGE [39]. Modeling the structure of the atmosphere with the weather research and forecasting (WRF) code (see Figs 5 and 8 in [40]) allowed us to confirm the charge structure of the lower atmosphere, which supports emerging of an electron accelerator.

Among other discoveries made on Aragats are registration of the atmospheric neutrons observed during thunderstorms, which proved to originate from the photonuclear reactions of the RREA gamma rays [41,42], the finding of the Radon circulation effect [43], uncovering of muon stopping effect [44], and estimation of the largest electric voltage (potential difference) at mountain peaks [45], observation of transient luminous events (TLEs) in the lower atmosphere [46].

In the following sections, we will present the methodology of TGE selection with statistical distributions of some important TGE parameters and highlight the most important results of TGE physics.

### 3. TGE selection criteria and statistical distributions of TGE parameters

From 2013 most particle detectors and spectrometers, field meters, and weather stations were operational providing 24/7 monitoring of almost all species of secondary cosmic rays, NSEF, and a variety of meteorological parameters.

In Fig. 1 we show the daily averaged time series of the NSEF (black), and outside temperature (blue) for 11 years from 2011 to 2022. We can see the anti-correlation of time series, which is obviously due to the scarcity of thunderstorms at Aragats in the winter months. The outside temperature is varying within (-25 - +20C) and NSEF (-40 – +40) kV/m, although the peak values corresponding to the lightning flashes can be larger.

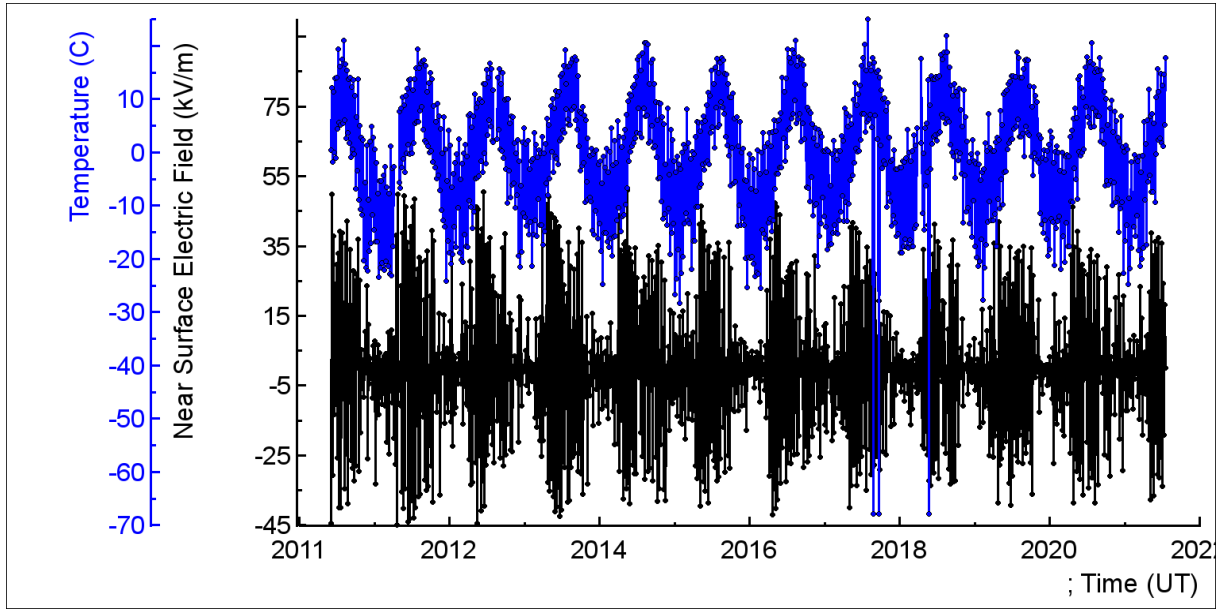

**Figure 1. Daily time series of the NSEF (electric mill EFM-100 by BOLTEK firm, black), and outside temperature (DAVIS weather station, blue)**

To identify the increase of the signal from a particle detector as a TGE event we check several detectors which are most stable during the considered time period. The measurement of TGE amplitude and its enhancements (in percent) and the statistical significance of the enhancement (in numbers of standard deviations from the mean value - $N\sigma$) are calculated relative to the pre-storm value of the mean count rate and its variance. No less than 3 different particle detectors, 2 of which have a threshold above 4 MeV, were used for the identification of TGE events. Thus, an enhancement is not considered valid unless confirmed by at least three detectors with significances not less than $3\sigma$. Afterward, the NSEF value is checked, the absolute value of which should exceed 5 kV/m.

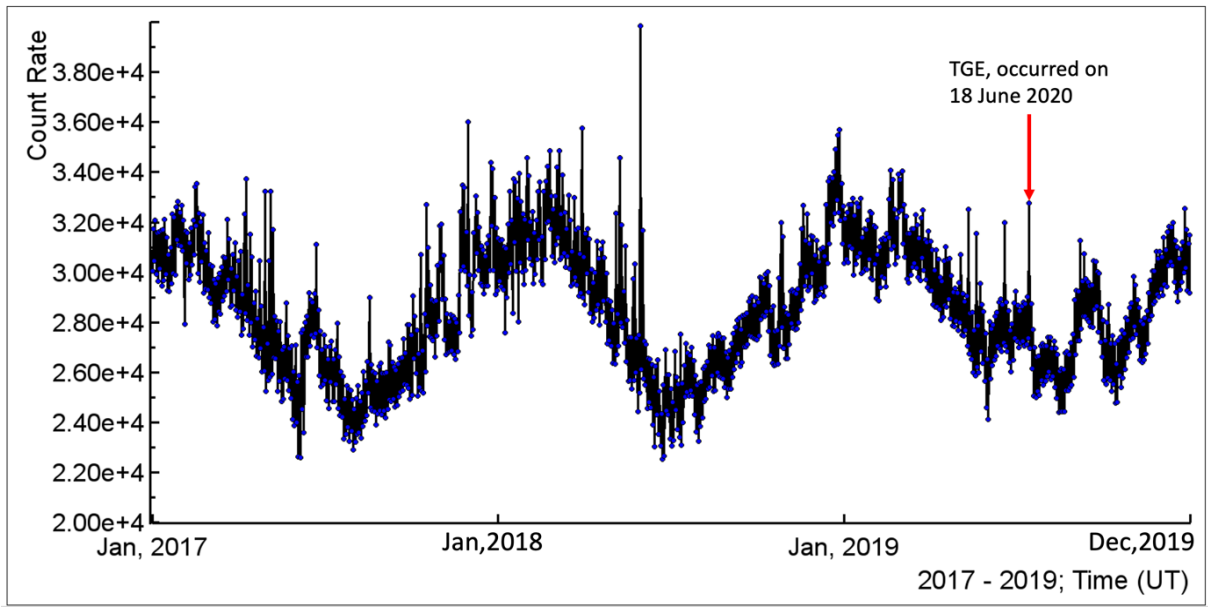

**Figure 2. 1-minute time series of count rates of 3 cm thick scintillator of the STAND3 detector, energy threshold 4 MeV used for TGE identification (covering 3 years' time span). By the arrow, we show one of the registered TGEs, its zoomed version will appear in the next picture.**

In Fig.2 we present the time series of count rates of the upper 3-sm thick scintillator of the STAND3 detector used for the TGE identification. In the Figure, we can see long-term periodicities connected with seasonal variations of the registered count rate and short outbursts due to abrupt enhancement of particle fluxes, which occurred during thunderstorms (TGEs). Each TGE should be confirmed by at least 3 independent detectors and by high absolute values of NSEF.

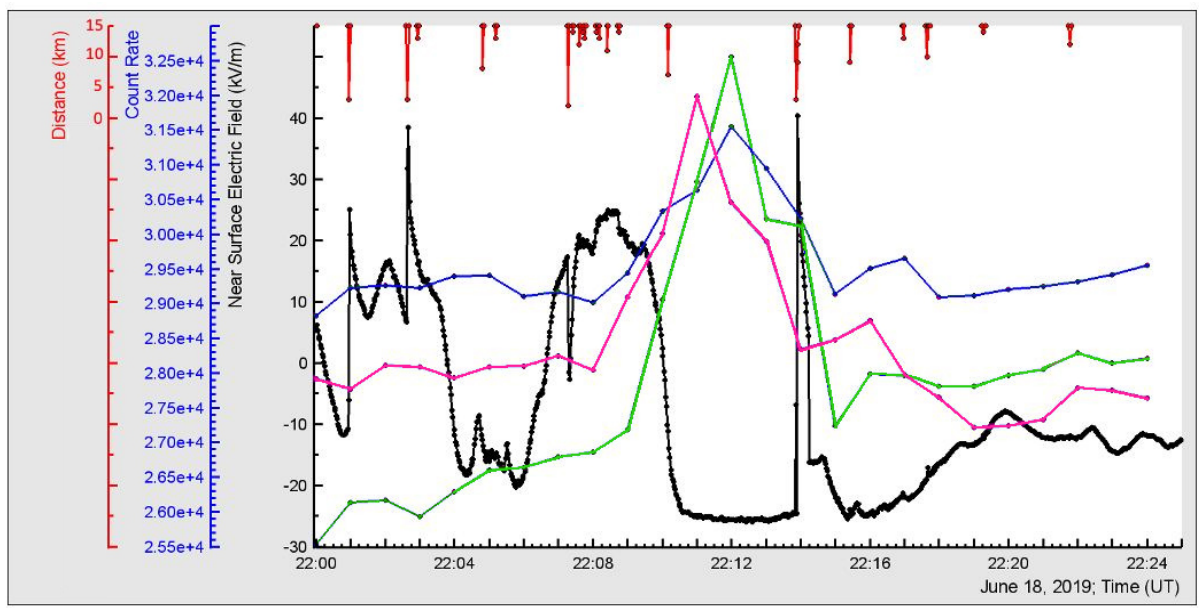

**Figure 3. Time series of scintillator counters: SEVAN (blue), STAND1 (green), and STAND3 magenta), see details in Table 1; disturbances of NSEF (black), and distances to lightning flash (red).**

In Fig. 3 we show 1 minute time series of count rates of three scintillation particle detectors with an area of 1 m$^2$, namely SEVAN (5 cm thick), STAND1 (1 cm thick), and STAND3 (3 cm thick). Also, we show the disturbances of NSEF and the distance to lightning flashes that occurred during TGE. TGE started at positive NSEF reaching 25 kV/m, the TGE maximum was reached during prolonged negative NSEF, being approximately constant for during ≈5 minutes and reaching – 25 kV/m. At 22:13:58 nearby lightning flash (distance to the flash 2.5 km) interrupted TGE.

In Table 1, we present the enhancements and significances of the TGE measured by these 3 detectors and their energy thresholds.

**Table 1. Energy thresholds, enhancements in percent, and significance of the peak values of the particle detectors participating in selecting TGE candidates**

|  | SEVAN | STAND1 | STAND3 |
|---|---|---|---|
| Energy threshold (MeV) | 7 | 0.8 | 5 |
| Enhancement (%) | 9 | 31 | 15 |
| Significance (Nσ) | 12 | 45 | 25 |

As seen in Fig. 3 and Table 1, all 3 detectors show a very high significance of TGE detection, and the chance probability of the peak to be a background fluctuation is negligible. NSEF reaches an absolute value of 25 kV/m, being positive at the beginning of TGE and turning to a deep negative at the maximum of TGE, and, finally, it returns to a positive value, although lower than before. Thus, this TGE satisfies all criteria and enters the database of TGE events measured on Aragats.

The yearly distribution of 564 TGE events registered from 2008 to 2022, and selected according to the criteria explained above, is presented in Fig 4a. On average, around 38 TGE events were registered yearly, however, the declining trend is apparent in 2018-2021 with 2019 registering the lowest TGE activity of 12 years – only 18 TGEs, being more than 4 times less than in 2010. This trend changed in 2021, for which the number of detected TGEs reached 36.

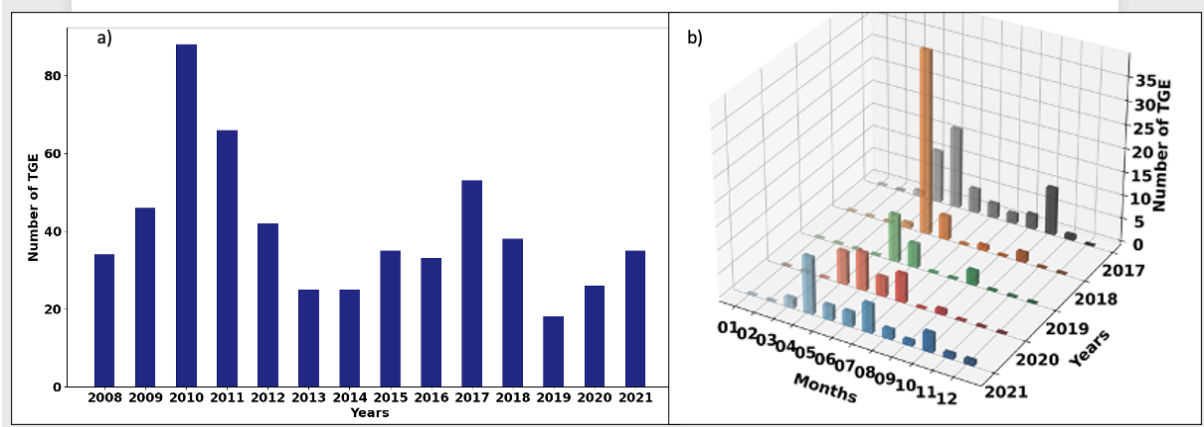

**Figure 4. TGE yearly and monthly statistics.**

In the monthly distribution of TGE events registered during the last 5 years and presented in Fig.4b we can see that most frequently, the TGEs are observed in May. However, a trend is noticed in moving the month of maximum TGEs to late months: in 2020 to June, in 2021 – to July.

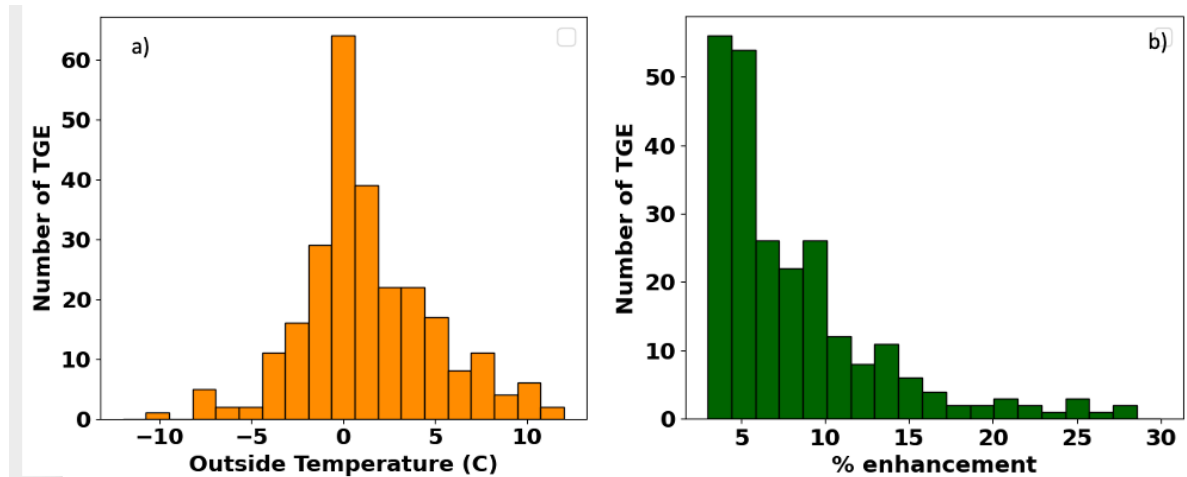

**Figure 5. a) The distribution of outside temperatures during TGEs; b) distribution of TGE significances by 3 cm thick plastic scintillator of STAND3 detector.**

In Fig. 5 we show important characteristics of a TGE subsample (a total of 288 TGEs registered from 2013 to 2021). The distribution of other TGE parameters can be obtained from the dataset posted in the Mendeley depository [47].

The distribution of temperatures shown in Fig. 5a peaked at 0° C, and the variance of the distribution is 2° C. Thus, the most probable temperature for the TGE occurrence lies in the interval from -2° to +2 C°. In Fig.5b, we show the distribution of TGE significances measured by the 3 cm thick scintillator of the STAND3 detector, the large significance values are very rare, only 4 of them are larger than 30σ (not shown in Fig.5b to be reported elsewhere).

### 4. Evidence for occurrence of TGEs prior to lightning flashes

After installing on Aragats a network of the EFM-100 electric field sensors in 2011, we registered numerous TGEs that were terminated by the lightning flash.

In two previous papers [48] and [49] published in 2017 and 2020 respectively, we examined small subsamples of TGE events terminated by lightning flash (24 events in 2017 and 13 events in 2020), and developed a methodology for lightning type identification. In the present research, we enlarged the statistics of the TGE events terminated by lightning flash (now 165 TGEs), identified lightning type, and estimate the distance to the lightning flash for every TGE. The data of 165 TGEs can be found in the Mendeley dataset [47]. The dataset is organized as a table with eleven columns, and 165 rows, that contain hyperlinks to the figures for each TGE event, demonstrating the abrupt termination of the TGE at the time of lightning discharge. This subsample of TGE events is of particular importance because it clearly indicates the association of TGEs with the electric field produced by thunderclouds.

An example of TGE terminated by lightning flash is presented in Fig. 6, where we show the time series of count rates of a scintillation detector (panel a) and the NSEF measured by two field mills at Aragats and Nor Amberd stations which are separated by 12.8 km (panels b and c). As seen from the figure, an abrupt decrease of particle count rate by ~15% is observed at the time of fast changes of the NSEF produced by lightning.

The polarity of a larger electrostatic field change (corresponding to the closer station) measured at Aragats station is negative (panel b), whereas the smaller field change in Nor Amberd is positive (panel c). The observed polarity reversal of electrostatic field change with distance indicates that this lightning discharge has partially destroyed a dipole, and this dipole was negative (negative charge above positive). In this paper, we use the atmospheric electricity sign convention, according to which the downward directed electric field or field change vector is positive. This lightning discharge was identified as an inverted intracloud (IC) flash followed by a negative cloud-to-ground (-CG) flash.

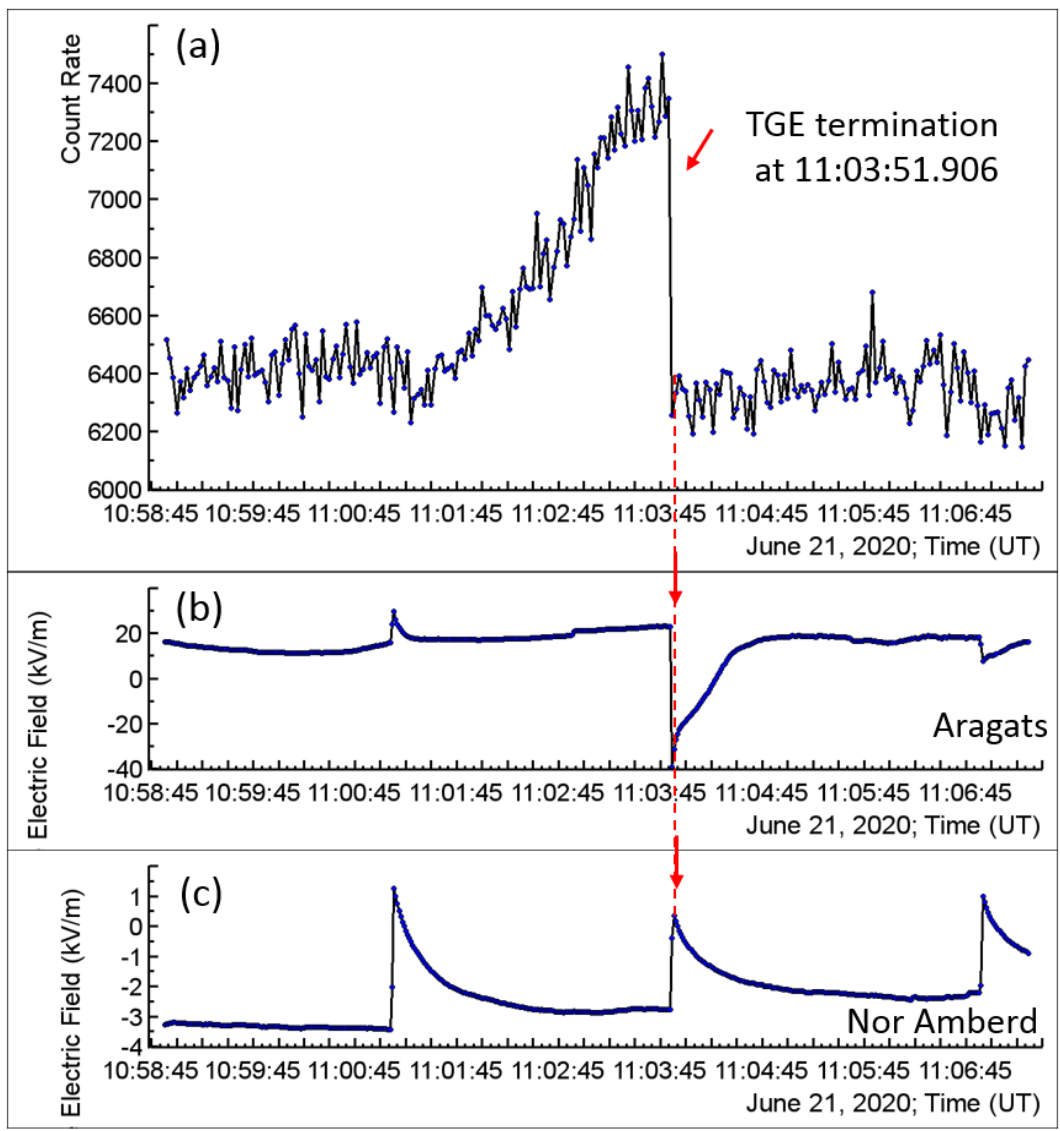

**Figure 6. TGE terminated by a lightning flash that occurred at 11:03:51.906 on June 21, 2020. The upper panel (a) shows the count rate of the scintillation detector. Panels (b) and (c) show the NSEF measured at Aragats and Nor Amberd stations, respectively.**

More than 70% of TGEs accompanied by nearby lightning flashes which occurred within 10 km from the detector site were abruptly terminated. 57 % of TGE events occurred during the positive NSEF, 38% - during negative, and for a small fraction of observed events (5%) the polarity varied from negative to positive or inversely. Assuming the classical tripole model of cloud charge structure, we suppose that the positive polarity of the NSEF is due to the presence of a large LPCR located low above the detector site. This is supported by the observation that in most cases of positive NSEF the TGE is terminated by an inverted intracloud (IC) flash (which occurs between the LPCR and the mid-level negative charge region), or hybrid lightning when an inverted IC flash is followed by a negative cloud-to-ground (-CG) flash.

If we consider far cloud-to-ground (-CG) lightning flashes, we can assume that the most intense atmospheric electric field was at the location where –CG struck. Therefore, this electric field initiates very intense TGE, which in turn provides an easy path for the lightning leader to quench the lower dipole. A strong electric field initiates intense electron-gamma ray avalanches when particle fluxes reach the earth's surface, and, as well, – lightning flashes when RREA electrons possibly made enough ionization to make easier the path for a lightning leader [50]. We yet have no elaborated model of the lightning initiation by RREA; therefore, our claim is that RREA is a precursor of lightning flash. This rather strong statement is supported by the research of other groups.

The group from Langmuir Laboratory in central New Mexico after examining of 23 thunderstorm electric field soundings suggest that lightning may occur whenever the electric field exceeds the critical field [51]. The same group observed during balloon flights on 3 July 1999 the maximal field of 1.86 kV/cm (130% of the threshold for a runaway process) at 5.77 km altitude just before nearby lightning flashes [52]. Authors conclude that RREA avalanches have limited the magnitude of the electric field inside storms and initiated lightning flashes.

Terrestrial gamma-ray flashes (TGFs) comprise very few high-energy photons from RREAs unleashed in the upper dipole and occasionally reach orbiting gamma observatories. They are associated with severe thunderstorms in the equatorial regions, however, due to the large distance from the source to spectrometers, located on the satellites, different scenarios of relative time-sequence of TGF and lightning are reported. Recently, the Atmosphere-Space Interactions Monitor (ASIM) consisting of X- and gamma-ray detectors, optical photometers, and cameras clarified the temporal relation between TGFs and lightning discharges. The authors of [53] conclude that TGFs precede lightning flashes.

5. **Revealing the electron flux in the TGE**

Facilities of Aragats research station were tuned to register enhanced particle fluxes during thunderstorms starting from 2008, and we can outline September 19, 2009, as a start of research of high-energy physics in the atmosphere (HEPA) on Aragats. The electric field meters and weather stations were not installed yet, however, the event that occurred in 2009 was so outstanding and convincing that we decided to enlarge Aragats facilities to encompass HEPA physics entirely. We already described this event in [54]. However, only in 2018, after performing multiple simulations of particle transport through the atmosphere and ASNT detector, and developing the methodology of solving inverse problems of cosmic rays, did we recognize that it is possible to retrieve the energy spectra of electrons and gamma rays separately.

Registration of electrons by the ground-based detectors is a direct proof of the RREA origin of the enhanced particle fluxes observed on the earth's surface. Only a very special configuration of the atmospheric electric field permits the RREA electrons to reach the spectrometers.

The spectrometer, that provides recovering the electron and gamma ray spectra, is the Aragats solar neutron telescope, shown in Fig.7 (see details of ASNT operation in [30]).

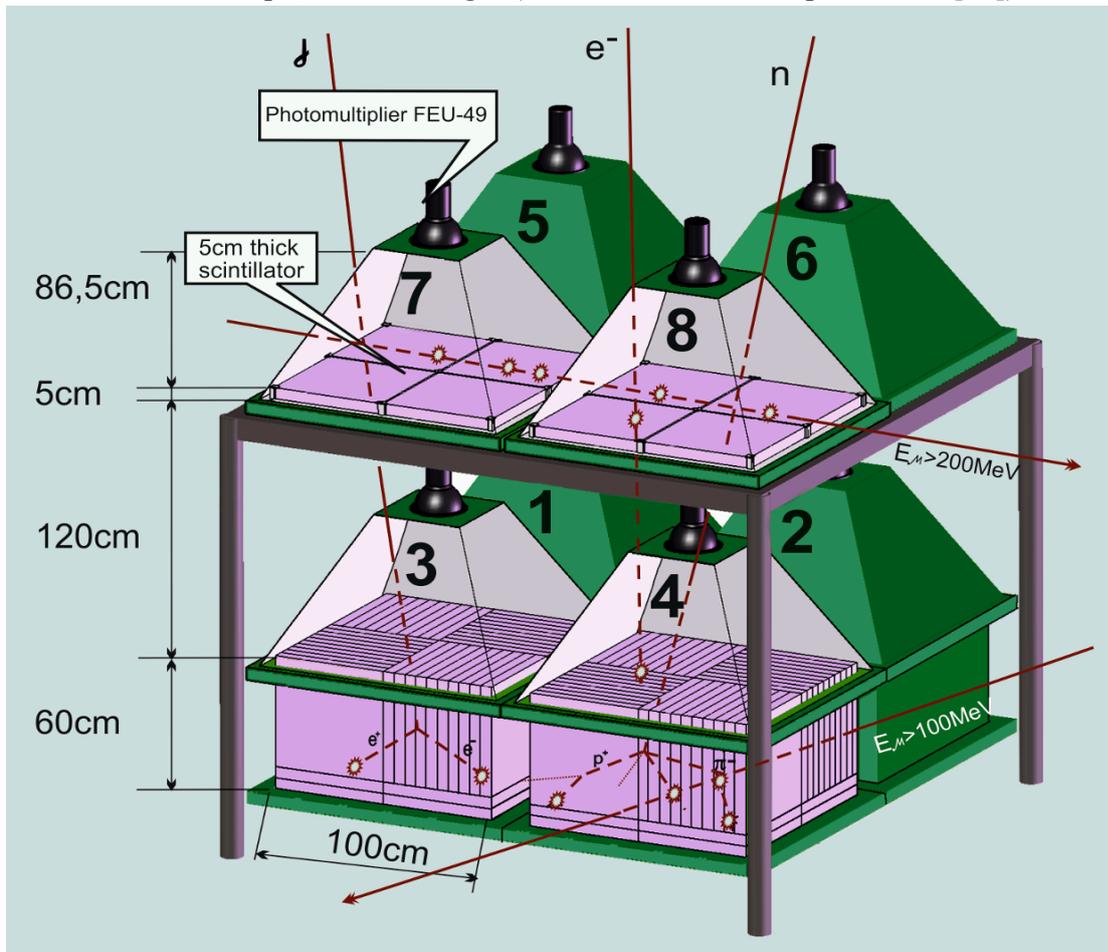

**Figure 7. Assembly of ASNT with a schematic view of the different particles traversing the detector. TGE electrons are registered by both upper and lower layers; gamma rays and neutrons – by invoking the veto option (no signals from the upper scintillators).**

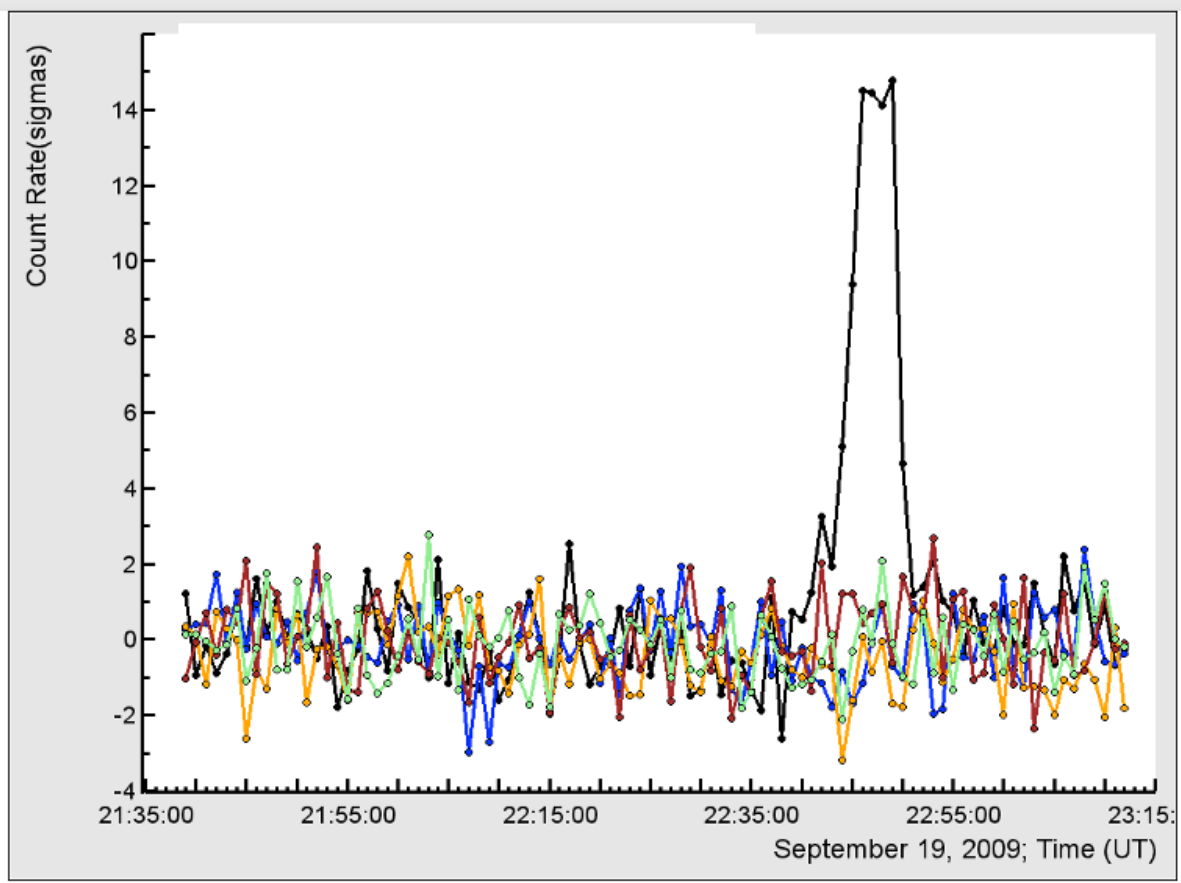

**Figure 8. Time series of the count rates of particles arriving from different directions, black – vertical direction (1-5,2-6,3-7,4-8), blue – direction connected 1-6 and 3-8 scintillators; red – direction 1-7 and 2-8; green – direction 2-5 and 4-7; yellow – direction 3-5 and 4-6.**

In Fig. 8, we can see the time series of count rates corresponding to the different inclinations of the incident particles. ASNT consists of four independent modules and all coincidences in modules belonging to upper and lower layers are counted separately, which allows one to discern the particle fluxes according to incidence angles. In Fig. 8 we show the count rates of particles coming within the near-vertical direction in the cone of 0º - 22º (black curve), and within the zenith angle of 22º - 58º from four different azimuth angles of 0, 90, 180, and 270 degrees (colored curves). As we see in Fig. 7b, only the vertical direction demonstrates a large peak, corresponding to the vertically arriving TGE particles because electrons are accelerated by a vertical electric field.

In Fig. 9a and 9c, we show the time series of count rates of two ASNT coincidences, namely the "01" – signal only in the lower layer of the ASNT spectrometer (60 cm thick scintillator), and "11" – signals in both layers of the spectrometer (also in the upper layer – 5 cm thick scintillator). We show the time series in units of a percent of the enhancement relative to mean values of the count rates at fair weather, the mean and variance of the time series are measured just before the TGE. The detection efficiency of a 5-cm scintillator is above 95% for electrons and 5-7% for gamma rays; for the 60-cm scintillator, the detection efficiency is above 95 % for electrons and 40-70 % for gamma rays. Thus, particles registered by "01" coincidence are mostly gamma rays, and by "11" coincidence – electrons. As we can in Fig. 9a, during the minute 2:03-2:04 ASNT registered a significant enhancement (reaching 5$\sigma$) of the "11" coincidence, which allows one to recover electron energy spectra. In Fig. 9b we show the recovered energy spectra of both species of TGE. The ratio of the electrons-to-gamma rays is obtained by the integration of the differential energy spectra shown in Fig. 9b from 10 MeV. The ratio reaches 13%, larger than the "11"/"01" ratio due to the much larger attenuation of the electrons compared to gamma rays in the upper scintillator of the ASNT spectrometer. To reach the lower scintillator, electrons cross the 5 cm thick scintillators and metallic tiles of the detector housings. The fraction of electrons can be rather large if the strong electric field extends almost to the ground, see in Fig. 9c the exceptional TGE occurred on 19 September 2009.

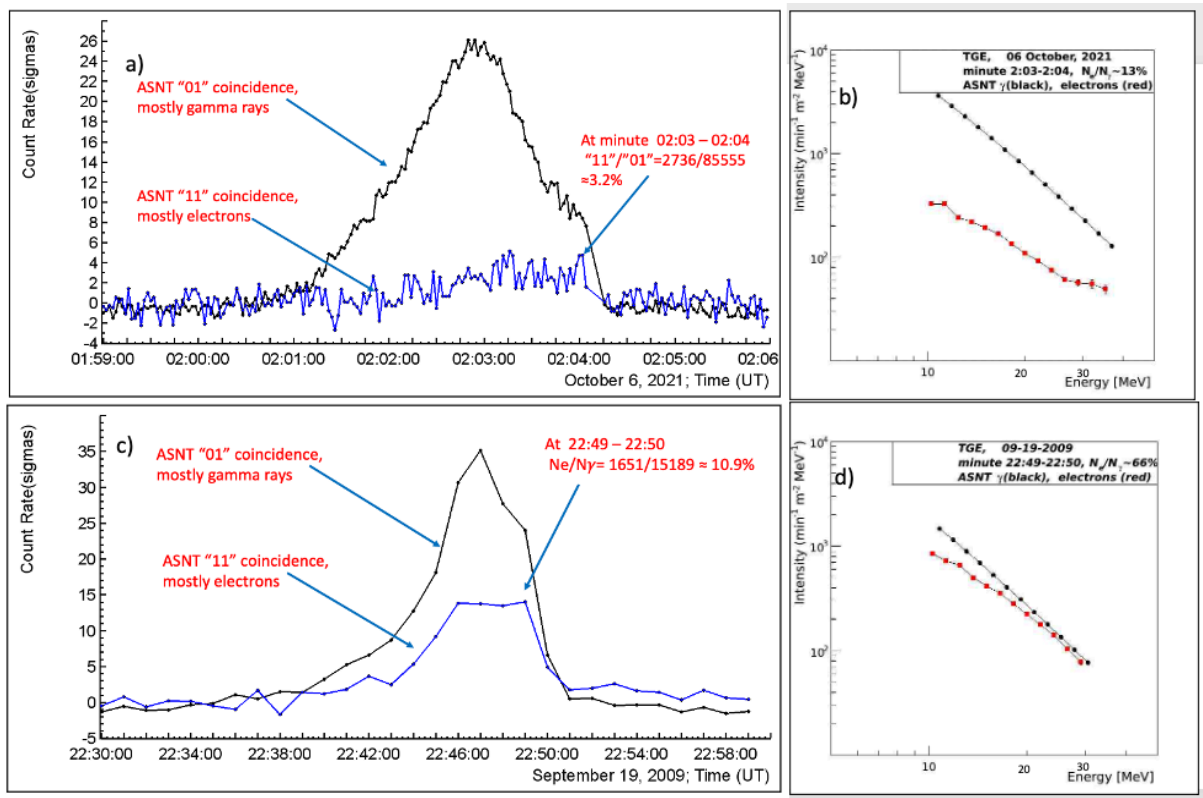

**Figure 9.** a) 2-s time series of count rates of coincidences registered by ASNT spectrometer shown in percent of the count rates measured at fair weather before TGE; b) differential energy spectra of TGE species recovered from energy release histograms; c) 1-min time series of count rates of coincidences registered by ASNT spectrometer; d) differential energy spectra of TGE species recovered from energy release histograms.

In Fig. 9a and 9c, we show the ASNT spectrometer's "01" and "11" coincidences for 1-minute and 2s time series of count rates recorded in 2009 and 2021 years (in 2009 the ASNT electronics sampling time was 1 minute). In Figures 9b and 9d, we show recovered differential energy spectra of electrons and gamma rays of these TGEs (see details of spectrum recovery method in [30]). The intensity of the electron flux measured on 6 October 2021 is significantly lower than the gamma ray intensity (Figs 9b). In contrast, as we can see in Fig. 9d, the intensities of electron and gamma ray spectra are almost equal for the TGE measured on 19 September 2009, and the electron to gamma ray ratio reaches 66% at the minute 22:49-22:50. It can happen only if a strong accelerating electric field is very low above the ground (we estimate the height to be 25-50 m) and an electron avalanche covers a sizable area on the ground. Additional evidence of a very low location of the strong electric field on 19 September 2009 is shown in Fig.10.

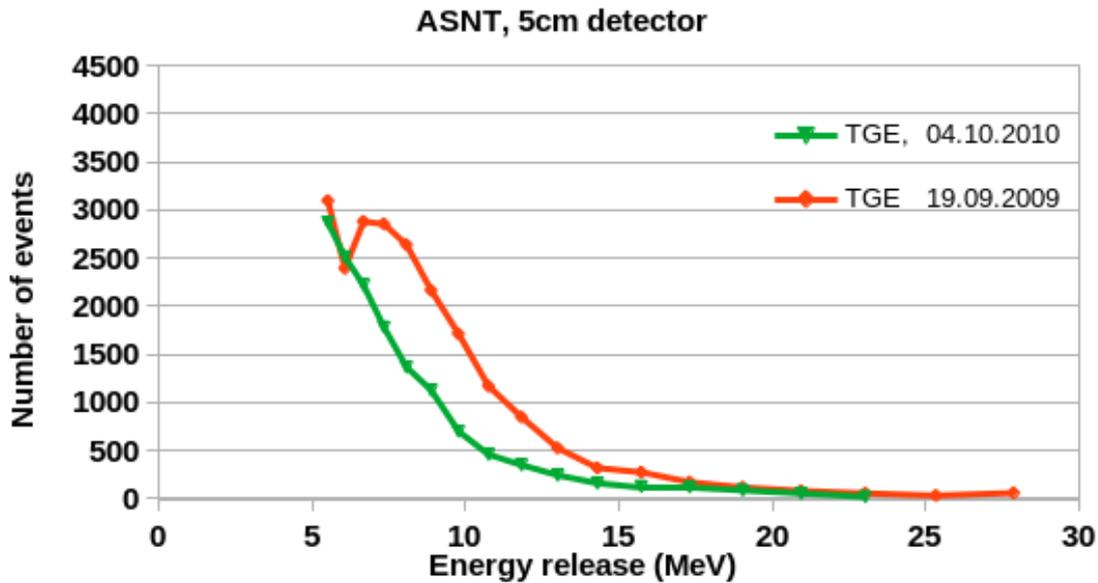

**Figure 10. Energy release histograms in the 5 cm thick scintillator of the ASNT detector measured during 2 TGES, on 19 September 2009 and 4 October 2010.**

In Fig. 10 we demonstrate the energy release histograms in the 5 cm thick scintillator at minute 2:49 - 22:50 on 19 September 2022 (red curve), which peaked at 8-9 MeV, as electron losses in the scintillator are ≈ 1.8 MeV per centimeter. Thus, high-energy TGE electrons (with energies of 10 MeV and more) only can induce this peak in the energy release distribution. The energy release distribution caused by TGE electrons is smeared by energy releases of gamma rays occasionally giving energy release in the 5 cm thick scintillator (the efficiency to register gamma rays is 5-7%). However, the energy releases of gamma rays have an exponential shape and do not produce any peak in the histogram, as we can see in Fig. 10 (green curve), where we show energy releases measured during another large TGE observed on 4 October 2010. This TGE does not contain electrons (the accelerating electric field stopped high above the earth's surface and the electron flux attenuate before reaching the spectrometer) and we detect no peak in the TGE energy release histogram. Thus, if we have a large peak in the time series of "11" coincidence (Figs. 9a and 9c) and a peak in the energy releases in the 5 cm thick scintillator (red curve in Fig. 10), we can be sure that TGE contains sizeable electron share. A long tail in the energy release distribution can be explained by pair production by high-energy gamma rays in the scintillator or in the metal of the scintillator housing.

## 6. Conclusions

We explain the methodology for the TGE selection and for proving electron content in the TGE. We present the distributions of TGE parameters from the databases collected during more than a decade of 24/7 monitoring of particle fluxes, environmental parameters, and near-surface electric fields. We investigated in detail the subclass of TGE events abruptly interrupted by lightning discharges. Lightning flashes that interrupt the TGE abruptly reduce the accelerating electric field in the thundercloud and, as a result, cause a sharp drop in the flux intensity of the TGE particles. Characteristics of the observed TGE events and selected datasets are available from the database of cosmic ray division and in datasets allocated in the Mendeley depository [47]. Thus, not only measurements used in this paper, but all multivariate measurements from the hundreds of channels are available for the community. We present our measurements on user-friendly sites allowing further analysis by colleagues worldwide. We provide links to each event with all additional measurements. Thus, the reader will be provided with materials to continue the correlation analysis in the multivariate space of the measurements and come to new interesting physical results.
On the basis of the analysis of TGEs observed during 12 years we conclude that

- The origin of a TGE is the RREA developed in the atmospheric electric field above the ground. TGEs are observed on Aragats by the multiple independent detectors and by SEVAN detectors on the mountain peaks of several East European countries (Musala in Bulgaria, Lomnicky Stit in Slovakia, etc.).

- For some TGE events, a strong accelerating electric field can extend very low above the earth's surface (50-150 m); correspondingly high-energy electrons (E>10 MeV) reach the ground in sizable amounts;
- Quite frequently, the TGE occurred prior to lightning flashes.

**Acknowledgements**

We thank the staff of the Aragats Space Environmental Center for the uninterruptable operation of all particle detectors and field meters. The authors acknowledge the support of the Science Committee of the Republic of Armenia (research project № 21AG-1C012), in the modernization of the technical infrastructure of high-altitude stations.

**Data availability statement**: The data that support the findings of this study are openly available at the following URL[database]: http://adei.crd.yerphi.am/.